\documentclass[pre,twocolumn,amsmath,amssymb,nofootinbib,floatfix,superscriptaddress]{revtex4}

\usepackage{graphicx,bm}
\usepackage[normalem]{ulem} 

\makeatletter
\def\graphicscale{\twocolumn@sw{0.3}{0.4}}
\def\graphicthreescale{\twocolumn@sw{0.3}{0.4}}

\begin{document}

\title{Three-dimensional phase transitions in 
multiflavor lattice scalar SO($N_c$) gauge theories}

\author{Claudio Bonati} 
\affiliation{Dipartimento di Fisica dell'Universit\`a di Pisa 
       and INFN Largo Pontecorvo 3, I-56127 Pisa, Italy}

\author{Andrea Pelissetto}
\affiliation{Dipartimento di Fisica dell'Universit\`a di Roma Sapienza
        and INFN Sezione di Roma I, I-00185 Roma, Italy}

\author{Ettore Vicari} 
\affiliation{Dipartimento di Fisica dell'Universit\`a di Pisa
       and INFN Largo Pontecorvo 3, I-56127 Pisa, Italy}

\date{\today}

\begin{abstract}
We investigate the phase diagram and finite-temperature transitions of
three-dimensional scalar SO($N_c$) gauge theories with $N_f\ge 2$
scalar flavors.  These models are constructed starting from a
maximally O($N$)-symmetric multicomponent scalar model ($N=N_cN_f$),
whose symmetry is partially gauged to obtain an SO($N_c$) gauge
theory, with O($N_f$) or U($N_f)$ global symmetry for $N_c \ge 3$ or
$N_c=2$, respectively. These systems undergo finite-temperature
transitions, where the global symmetry is broken.  Their nature is
discussed using the Landau-Ginzburg-Wilson (LGW) approach, based on a
gauge-invariant order parameter, and the continuum scalar SO($N_c$)
gauge theory. The LGW approach predicts that the transition is of
first order for $N_f\ge 3$. For $N_f =2$ the transition is predicted
to be continuous: it belongs to the O(3) vector universality class for
$N_c=2$ and to the $XY$ universality class for any $N_c\ge 3$. We
perform numerical simulations for $N_c=3$ and $N_f=2,3$. The numerical
results are in agreement with the LGW predictions.
\end{abstract}

\maketitle


\section{Introduction}
\label{intro}

Global and local gauge symmetries play a crucial role in theories
describing fundamental interactions~\cite{Weinberg-book} and emerging
phenomena in condensed matter physics~\cite{Sachdev-19}.  Interacting
scalar fields with local gauge symmetries provide paradigmatic
examples for the Higgs mechanism at the basis of
superconductivity~\cite{Anderson-63} and of the Standard Model of the
fundamental interactions~\cite{SSBgauge}.  In condensed matter
physics, they may be relevant for systems with emerging nonabelian
gauge symmetries, see, e.g., Refs.~\cite{GASVW-18,SSST-19}.  The
interplay between global and local gauge symmetries turns out to be
crucial to determine their phase diagram, the nature and universality
classes (if the transition is continuous) of their thermal and quantum
transitions.

These issues have been recently investigated in multicomponent lattice
Abelian-Higgs models~\cite{PV-19-2,BPV-19-2d} and in multiflavor
lattice scalar models with SU($N_c$) gauge
symmetry~\cite{BPV-19,BPV-20-3d,SPSS-20,BPV-20-2d}.  For
three-dimensional (3D) systems, the nature of the phase transitions
turns out to be effectively described by Landau-Ginzburg-Wilson (LGW)
$\Phi^4$ theories based on a gauge-invariant order-parameter field,
that have the same global symmetry as the lattice model. The LGW
approach is expected to be effective when the gauge interactions are
short-ranged at the transition and can therefore be neglected in the
effective model that encodes the long-range modes.  In the opposite
case, when gauge correlations become critical as well, other theories
may be more appropriate, such as continuum gauge theories in which
gauge fields are explicitly present.

In this paper we return on this issue, to deepen our understanding of
the role that global and local nonabelian symmetries play in
determining the main features of the phase diagram and the nature of
the phase transitions.  For this purpose, we consider a multiflavor 3D
lattice scalar model characterized by an SO($N_c$) gauge symmetry and
an O($N_f$) global symmetry, using the standard Wilson
formulation~\cite{Wilson-74}. The model is defined starting from an
O($N$)-symmetric scalar model with $N=N_f N_c$.  The global O($N$)
symmetry is partially gauged, obtaining a nonabelian gauge model, in
which the fields belong to the coset $S^N$/SO($N_c$), where
$S^N=\hbox{SO}(N)/\hbox{SO}(N-1)$ is the $N$-dimensional sphere.

In this paper, we shall show that the phase diagrams of multiflavor
lattice SO($N_c$) gauge models present two phases, which can be
characterized by using a rank-two real order parameter, whose
condensation breaks the global symmetry.  To identify the nature of
the phase transition, which separates the two phases, we consider two
different field-theoretical approaches: the effective LGW theory,
based on a gauge-invariant order-parameter field, and the continuum
multiflavor scalar SO($N_c$) gauge theory with explicit nonabelian
gauge fields. Their predictions are compared with numerical Monte
Carlo (MC) results.  As it was the case for the multiflavor lattice
scalar chromodynamics characterized by an SU($N_c$) gauge
symmetry~\cite{BPV-19,BPV-20-3d} and for the multicomponent lattice
Abelian-Higgs model with U(1) gauge symmetry~\cite{PV-19-2}, a
detailed finite-size scaling (FSS) analysis of the numerical results
supports the LGW predictions.  We recall that an analogous LGW
approach was originally used to predict the nature of the
finite-temperature phase transition of hadronic matter in the limit of
massless quarks, implicitly assuming that the SU(3) gauge modes are
not critical~\cite{PW-84,BPV-03,PV-13}.

The paper is organized as follows. In Sec.~\ref{sec:model} the lattice
model is introduced, with a discussion of its global and local
symmetry. In Sec.~\ref{sec.III} we define the LGW $\Phi^4$ theory
appropriate for the model and the continuum scalar SO($N_c$) gauge
theory and discuss their predictions for the nature of the
transitions.  In Sec.~\ref{numres} we report MC results for $N_c=3$
and $N_f=2,3$, and the FSS analyses that we perform to ascertain the
nature of the phase transitions.  Finally, we summarize and draw our
conclusions in Sec.~\ref{conclu}.

\section{The lattice model}
\label{sec:model}

We consider a 3D lattice model defined in terms of $N_c\times N_f$
real matrix variables $\varphi^{af}_{\bm x}$ associated with each site
${\bm x}$ of a cubic lattice. We start from a maximally symmetric model with action
\begin{align}
&S_{\rm inv} =- \sum_{{\bm x},\mu} {\rm Tr} \,\varphi_{\bm x}^t
  \varphi_{{\bm x}+\hat\mu} + \sum_{\bm x} 
V( {\rm Tr} \varphi_{\bm x}^t \,\varphi_{\bm x})\, , 
\label{hiom}\\ &V(X) = r\, X + {1\over
    2} u\, X^2\,,
\label{potential}
\end{align} 
where the first sum is over the lattice links, the second one is over
the lattice sites, and $\hat{\mu} =\hat{1},\hat{2},\hat{3}$ are unit
vectors along the three lattice directions. In this paper we consider
unit-length variables satisfying
\begin{equation}
{\rm  Tr}\,\varphi_{\bm x}^t \varphi_{\bm x} = 1\, ,
\label{phi2eq1}
\end{equation}
so that the action is simply
\begin{eqnarray} 
\label{ullimit}
S_{\rm inv} = - \sum_{{\bm x},\mu} 
{\rm Tr} \,\varphi_{\bm x}^t \varphi_{{\bm x}+\hat\mu} \,.
\end{eqnarray} 
Formally, the model can be obtained setting $r = - u$, and taking the
limit $u\to\infty$ of the potential~(\ref{potential}).  Models with
actions (\ref{hiom}) and (\ref{ullimit}) are invariant under O($N$)
transformations with $N = N_c N_f$. This is immediately checked if we
express the matrices $\varphi_{\bm x}$ in terms of $N$-component real
vectors ${\bm S}_{\bm x}$. In the new variables we obtain the standard
action of the O($N$) nonlinear $\sigma$-model
\begin{eqnarray}
S_N = - \sum_{{\bm x},\mu} {\bm S}_{\bm x}\cdot {\bm S}_{{\bm
    x}+\hat\mu}\,, \qquad {\bm S}_{\bm x} \cdot {\bm S}_{\bm x}=1\,.
\label{nvectorm}
\end{eqnarray} 
We now proceed by gauging some of the degrees of freedom: we associate
an SO($N_c$) matrix $V_{{\bm x},\mu}$ with each lattice link and
extend the action (\ref{ullimit}) to ensure SO($N_c$) gauge
invariance. We also add a kinetic term for the gauge variables in the
Wilson form \cite{Wilson-74}. We thus obtain the model with action
\begin{equation}
\begin{aligned} 
S_g  & = - N_f \sum_{{\bm x},\mu} 
{\rm Tr} \left[ \varphi_{\bm x}^t \, V_{{\bm x},{\mu}}
\, \varphi_{{\bm x}+\hat{\mu}}\right] \\
& - {\gamma\over N_c} \sum_{{\bm x},\mu>\nu} {\rm Tr}\,
\left[ V_{{\bm x},{\mu}} \,V_{{\bm x}+\hat{\mu},{\nu}} 
\,V_{{\bm x}+\hat{\nu},{\mu}}^t  
\,V_{{\bm x},{\nu}}^t\right]
\,,
\end{aligned}
\label{hgauge}
\end{equation}
and partition function
\begin{equation}
Z = \sum_{\{\varphi,V\}} e^{- \beta S_g}\,. \label{partfun}
\end{equation}
Note that, for $\gamma \to\infty$, the product of the gauge fields
along a plaquette converges to one. This implies that  $V_{{\bm x},\mu} =
1$ modulo a gauge transformation. Therefore, in the $\gamma\to\infty$  limit 
we reobtain
the O($N$) invariant theory (\ref{ullimit}) we started from.  For any
value of $N_c$ and $N_f$, $S_g$ is invariant under the local gauge
transformation $\varphi_{\bm x}\to G_{\bm x} \varphi_{\bm x}$ and
$V_{\bm x,{\mu}}\to G_{\bm x} V_{\bm x,{\mu}} G_{\bm{x}+\hat{\mu}}^t$
with $G_{\bm x}\in $ SO($N_c$), and under the global transformation
$\varphi_{\bm x}\to \varphi_{\bm x} W$ and $V_{\bm x,{\mu}}\to V_{\bm
  x,{\mu}}$ with $W\in $ O($N_f$).

For $N_c=2$ the global symmetry is actually larger than O($N_f$).  We
write $V_{{\bm x},\mu} \in $ SO(2) as
\begin{equation}
V_{{\bm x},\mu} = 
  \begin{pmatrix} \cos\theta_{{\bm x},\mu} & \sin\theta_{{\bm x},\mu} \\
                 -\sin\theta_{{\bm x},\mu} & \cos\theta_{{\bm x},\mu} 
  \end{pmatrix} \,,
\end{equation}
we define a complex $N_f$-dimensional vector 
\begin{equation}
{z}^f_{\bm x} = \varphi^{1f}_{\bm x} + i \varphi^{2f}_{\bm x}\,,
\label{zphirel}
\end{equation}
which satisfies $\bar{\bm
    z}_{\bm x} \cdot {\bm z}_{\bm x}=1$ because of
  Eq.~(\ref{phi2eq1}), and the U(1) link variable $\lambda_{{\bm
    x},\mu}\equiv e^{i\theta_{{\bm x},\mu}}$.  In terms of the new
variables, the lattice action (\ref{hgauge}) becomes
\begin{eqnarray}
&&S_{\rm AH} = - \,\sum_{{\bm x}, \mu} {\rm Re} \,
  [\bar{\bm{z}}_{\bm x} \cdot \lambda_{{\bm x},\mu}\, {\bm z}_{{\bm
        x}+\hat\mu}]
\label{gllf}\\
&&\qquad -\gamma \sum_{{\bm x},\mu>\nu} 
{\rm Re}\,[
\lambda_{{\bm x},{\mu}} \,\lambda_{{\bm x}+\hat{\mu},{\nu}} 
\,\bar{\lambda}_{{\bm x}+\hat{\nu},{\mu}}  
  \,\bar{\lambda}_{{\bm x},{\nu}} ]\,.
\nonumber
\end{eqnarray}
This is the action of the $N_f$-component lattice Abelian-Higgs model,
which is invariant under local U(1) and global U$(N_f)$
transformations. There is therefore an enlargement of the global
symmetry of the model: the global symmetry group is U$(N_f)$ instead
of O($N_f$). The phase structure of the Abelian-Higgs model
(\ref{gllf}) has been studied in detail in
Ref.~\cite{PV-19-2}. Therefore, in this work we will focus on the
behavior for $N_c \ge 3$.

It is interesting to note that one can consider more general
Hamiltonians that have the same global and local invariance. Indeed,
one can start from a Hamiltonian in which the potential is any
O($N_f$)-invariant function of $\varphi^t \varphi$. For instance, if
we only consider quartic potentials in $\varphi_{\bm x}$, we can take
\begin{equation}
V_g(\varphi) = V(X) +  
v \, \{ {\rm Tr} [\varphi_{\bm x}^t \,\varphi_{\bm x}
\,\varphi_{\bm x}^t \,\varphi_{\bm x}]
- ({\rm Tr} [\varphi_{\bm x}^t \,\varphi_{\bm x}])^2\}
\,.
\label{bgdef} 
\end{equation} 
If we consider this class of more general Hamiltonians, there is no
enlargement of the symmetry from O($N_f$) to O($N$) in the limit
$\gamma\to\infty$, in which gauge degrees of freedom are
frozen. Moreover, for $N_c = 2$, the symmetry enlargement from
O$(N_f)$ to U($N_f$) does not occur.

Since the global symmetry group O($N_f$) corresponds to SO($N_f)\times
{\mathbb Z}_2$, there is the possibility of breaking separately the
two different groups. In this work we will focus on the breaking of
the SO($N_f$) subgroup, which, by analogy with our results for complex
U($N_f$) invariant gauge models \cite{BPV-20-3d}, is expected to be
the only one occurring in the model with action (\ref{hgauge}).
However, the ${\mathbb Z}_2$ symmetry may play a role in more general
models, for instance in those with action (\ref{bgdef}), in which the
breaking of both the ${\mathbb Z}_2$ and the SO($N_f$) subgroups may
occur.  Note that the presence of two possible symmetry breaking
patterns is related to the fact that the gauge symmetry group is
SO($N_c$). Had we considered an O($N_c$) gauge invariant model, we
would have only an SO($N_f$) global invariance.

The natural order parameter for the breaking of the SO($N_f$) global
symmetry group is the gauge-invariant real traceless and symmetric
bilinear operator
\begin{equation}
Q^{fg}_{\bm x} = \sum_a \varphi^{af}_{\bm x} \varphi^{ag}_{\bm x} - 
     {\delta^{fg}\over N_f}\, ,
\label{qdef}
\end{equation}
which is a rank-2 operator with respect to the global O($N_f$)
symmetry group. As we shall show, the phase diagram of the model
($N_c\ge 3$) presents two different phases, separated by a
phase-transition line associated with the condensation of the bilinear
$Q$.

We finally mention that, for $N_f=1$, the phase diagram of the lattice
scalar SO($N_c$) gauge model (\ref{hgauge}) is expected to show only
one phase. This can be easily verified for $\gamma=0$. In this case
the $N_f=1$ model is trivial and cannot have any phase transition.

\section{Effective field theories} \label{sec.III}

\subsection{The LGW field theory}
\label{LGWth}

To characterize the finite-temperature transitions of scalar SO($N_c$)
gauge theories, we consider the LGW
approach~\cite{Landau-book,WK-74,Fisher-75,ZJ-book}.  We start by
considering an order parameter that breaks the global symmetry of the
model.  For $N_c \ge 3$, the global symmetry group is O($N_f$) and an
appropriate order parameter is the bilinear tensor $Q_{\bm x}$ defined
in Eq.~(\ref{qdef}).  The corresponding LGW theory is obtained by
considering a real symmetric traceless $N_f\times N_f$ matrix field
$\Phi({\bm x})$, which represents a coarse-grained version of $Q_{\bm
  x}$. The Lagrangian is
  \begin{eqnarray}
{\cal L}_{\rm LGW} &=& \hbox{Tr }\partial_\mu \Phi \partial_\mu \Phi + 
    r\, \hbox{Tr } \Phi^2  + \,u_3\, \hbox{Tr } \Phi^3 \qquad
\nonumber\\
&& 
+ \, u_{41} \,\hbox{Tr } \Phi^4 + 
      u_{42}\, (\hbox{Tr } \Phi^2)^2\,,
\label{SLGW} 
\end{eqnarray}
where the potential is the most
general O($N_f$)-invariant fourth-order polynomial in the field.
The Lagrangian (\ref{SLGW}) is invariant under the global
transformations 
\begin{equation}
\Phi \to W \Phi W^t\,,\qquad W\in{\rm O}(N_f)\,.
\label{ongtra}
\end{equation}

The renormalization-group (RG) flow of model (\ref{SLGW}) has been
already discussed in Ref.~\cite{PTV-18}.  For $N_f=2$ the cubic term
vanishes and the two quartic terms are proportional, so that we obtain
the two-component vector $\Phi^4$ action.  Therefore, for $N_f=2$ the
system may undergo a continuous transition in the $XY$ universality
class. For $N_f > 2$ the cubic term is generically present.  Assuming
that the usual mean-field arguments, valid close to four dimensions,
apply also to the three-dimensional case, only first-order transitions
are expected.  A continuous transition is only possible if the
Hamiltonian parameters are tuned or an additional symmetry is present,
so that the cubic term vanishes. If this occurs, we obtain the LGW
model discussed in Ref.~\cite{PTV-18}, in the context of the
antiferromagnetic RP$^{N_f-1}$ model.  In particular, for $N_f = 3$,
the LGW theory is equivalent to that of the O(5) vector model
\cite{PTV-18,FMSTV-05}, so that continuous transitions in the O(5)
universality class may occur.

The previous conclusions also hold for $N_c = 2$ for generic actions
with SO($N_f$) global symmetry, for instance for the action
(\ref{bgdef}). On the other hand, for our model (\ref{hgauge}) the
previous results do not hold for $N_c = 2$, because of the symmetry
enlargement to U($N_f$). In this case the LGW field is a Hermitean
traceless $N_f\times N_f$ matrix field~\cite{DPV-15,PV-19} with a
Lagrangian that is the analogue of the one considered here,
Eq.~(\ref{SLGW}). Its RG flow predicts
that~\cite{PV-19-2} the transition can be continuous for $N_f=2$, in
the O(3) vector universality class, while it is of first order for
$N_f\ge 3$.

The above-reported discussion applies to any model in which the global
symmetry group is SO($N_f$) and the order parameter is a real
operator that transforms as a rank-two tensor under SO($N_f$)
transformations. Therefore, the results apply to other scalar models
and, in particular, to scalar SU(2) gauge theories with scalar fields
in the adjoint representation, which have been recently considered to
describe the critical behavior of cuprate superconductors for optimal
doping~\cite{SSST-19,SPSS-20}.  In these theories the fundamental
fields are $N_h$ Higgs fields transforming under the adjoint
representation of SU(2), i.e. $\hat\phi^f_{\bm x} \equiv \sum_{a=1}^3
\phi_{\bm x}^{af} \tau^a$, where $\tau^a\equiv \sigma^a/2$, $\sigma^a$
are the Pauli matrices and $f=1,...,N_h$.  In the fixed-length limit
${\rm Tr}[\phi_{\bm x}^t\phi_{\bm x}] = 1$, the lattice action
is~\cite{Laine-95,SPSS-20}
\begin{equation}
\begin{aligned} 
S_h  & = - \sum_{{\bm x},\mu,f} 
{\rm Tr} \left[ \hat\phi^f_{\bm x} U_{{\bm x},{\mu}} 
\hat\phi^f_{{\bm x}+\hat{\mu}} U_{{\bm x},{\mu}}^\dagger \right] \\
& - {\gamma\over 2} \sum_{{\bm x},\mu>\nu} {\rm Tr}\,
\left[ U_{{\bm x},{\mu}} \,U_{{\bm x}+\hat{\mu},{\nu}} 
\,U_{{\bm x}+\hat{\nu},{\mu}}^t  
\,U_{{\bm x},{\nu}}^t\right]\\
&+ u
\sum_{\bm x}{\rm Tr} [\phi_{\bm x}^t \phi_{\bm x}\phi_{\bm x}^t \phi_{\bm x}]\,,
\end{aligned}
\label{hsu2gauge}
\end{equation}
where $U_{{\bm x},{\mu}}$ are SU(2) link variables.  For any $N_h$,
the action $S_h$ is invariant under the local SU(2) gauge
transformation $\hat\phi_{\bm x}^f\to G_{\bm x} \hat\phi_{\bm x}^f
G_{\bm x}^\dagger$ and $U_{\bm x,{\mu}}\to G_{\bm x} U_{\bm x,{\mu}}
G_{\bm{x}+\hat{\mu}}^\dagger$ with $G_{\bm x}\in $ SU(2), and under
the global transformation $\phi_{\bm x}\to \phi_{\bm x} W$ with $W\in
\,$O($N_h$).  The appropriate order parameter is again a
gauge-invariant operator which transforms as a rank-two traceless real
tensor with respect to the global O($N_h$) symmetry,
\begin{equation}
{\cal Q}^{fg}_{\bm x} = 
\sum_a \phi^{af}_{\bm x} \phi^{ag}_{\bm x} - 
     {\delta^{fg}\over N_h}\,.
\label{qsdef}
\end{equation}
The corresponding LGW action is again Eq.~(\ref{SLGW}).  Therefore,
for $N_h=2$ transitions associated with the breaking of the O($N_h$)
symmetry may be continuous in the $XY$ universality class.  For $N_h
\ge 3$ only first-order transitions are possible.

\subsection{The continuum scalar SO($N_c$) gauge theory}
\label{cogauth}

The continuum scalar SO($N_c$) gauge theory provides another effective
theory for the lattice model (\ref{hgauge}).  Its Lagrangian is
obtained by considering all monomials up to dimension four, which can
be constructed using the scalar field $\Phi_{af}$ (with $a=1,...,N_c$
and $f=1,...,N_f$). Gauge invariance is obtained as usual, by adding a
gauge field $A_{\mu,ab}\equiv {\cal A}_{\mu}^k T_{ab}^k$, where the
matrices $T^k$ are the generators of the SO($N_c$) gauge algebra.  The
Lagrangian reads~\cite{Hikami-80,PRV-01}
\begin{eqnarray}
&&{\cal L} = {1\over 4}F_{\mu\nu}^2 + 
{1\over 2}\sum_{af\mu} (\partial_\mu \Phi_{af} + g_0 A_{\mu,ab} \Phi_{bf})^2
\label{cogau}\\
&&\quad + {1\over 2} r_0 \sum_{af} \Phi_{af}^2  
+ {1\over 4!} u_0 ( \sum_{af} \Phi_{af}^2 )^2  \nonumber\\
&&\quad
+ {1\over 4!} v_0 \left [ \sum_{abf} \Phi_{af}^2 \Phi_{bf}^2 -    
( \sum_{af} \Phi_{af}^2 )^2 \right ] \,,
\nonumber 
\end{eqnarray}
where $F_{\mu\nu}$ is the non-Abelian field strength associated with
the gauge field $A_{\mu,ab}$.

To determine the nature of the transitions
described by the continuum SO($N_c$) gauge theory  (\ref{cogau}),
one studies the RG flow
determined by the $\beta$ functions of the model.  In the
$\epsilon$-expansion framework, the one-loop $\overline{\rm MS}$
$\beta$ functions control the RG flow close to four dimensions.
Introducing the renormalized couplings $u$, $v$, and $\alpha = g^2$,
the one-loop $\beta$ functions are (see Ref.~\cite{Hikami-80} for the
exact normalizations of the renormalized couplings) 
\begin{eqnarray}
&&\beta_u = -\epsilon u + {N_f N_c + 8\over 6} u^2   
+ {(N_f-1)(N_c-1)\over 6} (v^2 - 2 u v) 
\nonumber\\
&& \qquad -
{3\over 2}(N_c-1) u \alpha  + {9\over 8}(N_c-1)\alpha^2 \,,
\label{betas}\\
&&\beta_v = - \epsilon v +
{N_f+N_c - 8\over 6}v^2 + 2uv - {3\over 2}(N_c-1)v\alpha
\nonumber \\
&&\qquad  + {9\over 4}
(N_c-2)\alpha^2\,,
\nonumber\\
&&\beta_\alpha =
-\epsilon \alpha  + {N_f  - 22(N_c-2)\over 12}\alpha^2
\,,\nonumber
\end{eqnarray}
where $\epsilon\equiv 4-d$.  Note that for $N_c=2$ the $\beta$
functions $\beta_u$ and $\beta_\alpha$ for $v=0$ map exactly onto
those of the Abelian-Higgs model~\cite{HLM-74,PV-19-2}, after an
appropriate change of normalization of the couplings.  We recall that
the RG flow of the Abelian-Higgs theory has a stable fixed point only
for large $N_f$, i.e. $N_f>90+24\sqrt{15}+O(\epsilon)$.

The RG flow described by the $\beta$ functions (\ref{betas})
generally predicts first-order transitions, unless the number of
flavors is large. In particular, 
one can easily see that
the RG flow described by the $\beta$ functions (\ref{betas}) cannot
have stable fixed points for
 \begin{equation}
 N_f < 22(N_c-2)\,, 
 \label{nfb}
 \end{equation}
 for which the fixed points must necessarily have $\alpha=0$, at least
 sufficiently  close to
 four dimensions.  The fixed points with vanishing gauge coupling
 $\alpha=0$ are always unstable with respect to the gauge coupling,
 since their stability matrix $\Omega_{ij} = \partial \beta_i/\partial
 g_j$ has a a negative eigenvalue
\begin{equation}
\lambda_\alpha = 
\left. {\partial \beta_\alpha \over \partial \alpha} \right|_{\alpha=0}
= - \epsilon + O(\epsilon^2)\,.
\label{lambdares}
\end{equation}
A more careful analysis shows that, for any $N_c$, a nontrivial stable
fixed point (with nonzero values of all couplings) exists only for
sufficiently large $N_f$~\cite{Hikami-80,PRV-01}.  This result is also
confirmed by three-dimensional large-$N_f$ computations for fixed
$N_c$ ~\cite{Hikami-80}.  Therefore, continuous transitions are only
possible for a large number of components.

The above results contradict the LGW predictions.  For $N_f=2$ the
continuum theory predicts a first-order transition, while, according
to the LGW analysis, $XY$ continuous transitions are possible.  Vice
versa, for large $N_f$ continuous transitions are possible according
to the continuum theory, but not on the basis of the LGW analysis.  We
note that analogously contradictory results were obtained for the
Abelian-Higgs model~\cite{PV-19-2} and the scalar
chromodynamics~\cite{BPV-19}.

Note that, unlike the LGW theory (\ref{SLGW}), the RG flow of the
continuum scalar SO($N_c$) gauge theory (\ref{cogau}) presents an
unstable O($N$) fixed point with $N=N_f N_c$, which describes the
critical behavior of the lattice model (\ref{hgauge}) in the
$\gamma\to\infty$ limit.  This is located at
\begin{equation}
u = \epsilon \,{6\over N_fN_c+8}+O(\epsilon^2)\,,
\quad v=0\,,\quad \alpha=0\,.
\label{OMfp}
\end{equation}
One can easily show that this fixed point is always unstable, even in
the absence of gauge interactions. The perturbation associated with
the coupling $v$ is a spin-four perturbation at the O($N$) fixed
point, which is relevant for any $N\ge 3$~\cite{CPV-03,HV-11}.
Moreover, also the gauge perturbation associated with the coupling
$\alpha$ is relevant, as it is associated with a negative eigenvalue
$\lambda_\alpha = - \epsilon + O(\epsilon^2)$ of the stability matrix,
see Eq.~(\ref{lambdares}).
 
\section{Numerical results}
\label{numres}
 
In this section we report numerical results for the lattice scalar
SO(3) gauge theory with two and three flavors. We consider cubic
lattices of linear size $L$ with periodic boundary conditions.  To
update the gauge fields we use an overrelaxation algorithm implemented
\emph{\`a la} Cabibbo-Marinari \cite{Cabibbo:1982zn}, considering
three SO(2) subgroups of SO(3). We use a combination of
biased-Metropolis updates\footnote{In the biased-Metropolis algorithm,
  links are generated according to a Gaussian approximation of the
  action and then accepted or rejected by a Metropolis step
  \cite{Metropolis:1953am}; the acceptance ratio was larger than
  $90\%$ in all cases studied.} and microcanonical steps
\cite{Creutz:1987xi} in the ratio 1:5. For the update of the scalar
fields a combination of Metropolis and microcanonical updates is used,
with the Metropolis step tuned to have an acceptance rate of
approximately 30\%.

\subsection{Observables and analysis method}
\label{sec:obs}

We consider the energy density and the specific heat, defined as
\begin{eqnarray}\label{ecvdef}
E = -\frac{1}{3N_f V} \langle S_g \rangle\,,\quad
C_V =\frac{1}{9N_f^2V}\left( \langle S_g^2 \rangle 
- \langle S_g  \rangle^2\right)\,,\quad
\end{eqnarray}
where $V=L^3$. 
To study the breaking of the O($N_f$) flavor symmetry, we consider the
order parameter $Q_{\bm x}$ defined in Eq.~(\ref{qdef}).  Its
two-point correlation function is defined by
\begin{equation} \label{gxyp}
G({\bm x}-{\bm y}) = \langle {\rm Tr}\, Q_{\bm x} Q_{\bm y} \rangle\,,  
\end{equation}
where the translation invariance of the system has been explicitly
taken into account. We define the corresponding susceptibility $\chi$
and correlation length $\xi$ as
\begin{eqnarray}
\chi=\sum_{\bm x} G({\bm x}),\quad \xi^2 = \frac{1}{4
    \sin^2 (\pi/L)} \frac{\widetilde{G}({\bm 0}) - \widetilde{G}({\bm
      p}_m)}{\widetilde{G}({\bm p}_m)},\quad
\label{xidefpb}
\end{eqnarray}
where $\widetilde{G}({\bm p})=\sum_{{\bm x}} e^{i{\bm p}\cdot {\bm x}}
G({\bm x})$ is the Fourier transform of $G({\bm x})$ and ${\bm p}_m =
(2\pi/L,0,0)$.  

At continuous transitions, RG-invariant quantities, such as the Binder
parameter
\begin{equation}
U = \frac{\langle \mu_2^2\rangle}{\langle \mu_2 \rangle^2} \,, \qquad
\mu_2 = \frac{1}{V^2}  
\sum_{{\bm x},{\bm y}} {\rm Tr}\,Q_{\bm x} Q_{\bm y}\,,
\label{binderdef}
\end{equation}
and 
\begin{equation}\label{rxidef}
R_{\xi}=\xi/L\,,
\end{equation}
(which we generically denote by $R$), are expected to scale as
\cite{PV-02}
\begin{eqnarray}
R(\beta,L) = f_R(X) +  L^{-\omega} g_R(X) + \ldots \,, \label{scalbeh}
\end{eqnarray}
where $X = (\beta-\beta_c)L^{1/\nu}$ and next-to-leading scaling
corrections have been neglected. The function $f_R(X)$ is universal up
to a multiplicative rescaling of its argument, $\nu$ is the critical
exponent associated with the correlation length, and $\omega$ is the
exponent associated with the leading irrelevant operator.  In
particular, $U^*\equiv f_U(0)$ and $R_\xi^*\equiv f_{R_\xi}(0)$ are
universal, depending only on the boundary conditions and aspect ratio
of the lattice.  Since $R_\xi$ defined in Eq.~\eqref{rxidef} is an
increasing function of $\beta$, we can write
\begin{equation}\label{uvsrxi}
U(\beta,L) = F_U(R_\xi) + O(L^{-\omega})\,,
\end{equation}
where $F_U$ now depends on the universality class, boundary conditions
and lattice shape, without any nonuniversal multiplicative factor.
The scaling \eqref{uvsrxi} is particularly convenient to test
universality-class predictions, since it permits easy comparisons
between different models without requiring a tuning of nonuniversal
parameters.
 
In the following we will show that the critical behavior along the
phase transition line of two-flavor SO(3) gauge models belongs to the
$XY$ universality class, by verifying that the asymptotic FSS behavior
of $U$ versus $R_\xi$, see Eq.~(\ref{uvsrxi}), matches that obtained
for the $XY$ model.  On the other hand, for $N_f = 3$, we will show
that Eq.~(\ref{uvsrxi}) is not satisfied---the data of $U$ do not
scale on a single curve when plotted versus $R_\xi$. This can be taken
as evidence that the transition is of first order, a conclusion that
will be also confirmed by the two-peak structure of the distributions
of the energy.

\subsection{The two-flavor lattice SO(3) gauge model}
\label{twofla}

We now present numerical results for the two-flavor SO(3) gauge model
(\ref{hgauge}), showing that it undergoes continuous transitions in
the 3D $XY$ universality class, as predicted by the corresponding LGW
theory. Some accurate results for the universal quantities of the 3D
$XY$ universality class are reported in Table~\ref{tablexy}.

\begin{table}
\begin{tabular}{clllll}
\hline\hline
Ref. & 
\multicolumn{1}{c}{$\nu$} & 
\multicolumn{1}{c}{$\eta$} & 
\multicolumn{1}{c}{$\omega$} & 
\multicolumn{1}{c}{$R_\xi^*$} & 
\multicolumn{1}{c}{$U^*$} \\\hline
\cite{CHPV-06} & 0.6717(1) & 0.0381(2) & 0.785(20) & 0.5924(4) & 1.2432(2) \\
\cite{Hasenbusch-19} & 0.67169(7) & 0.03810(8) & 0.789(4) & 0.59238(7) & 1.24296(8) \\
\cite{CLLPSSV-19} & 0.67175(10) & 0.038176(44) & 0.794(8) & & \\
\hline\hline
\end{tabular}
\caption{Some estimates of the universal critical exponents for the 3D $XY$
  universality class, obtained from the analysis of high-temperature
  expansions supplemented by MC simulations~\cite{CHPV-06}, from MC
  simulations~\cite{Hasenbusch-19} and the conformal-bootstrap
  approach~\cite{CLLPSSV-19}.  See Ref.~\cite{PV-02} for earlier
  theoretical estimates and experimental results.  We report the
  correlation-length exponent $\nu$, the order-parameter exponent
  $\eta$, the exponent $\omega$ (associated with the leading scaling
  corrections), and the universal quantities $R_\xi^*$ and $U^*$
  (large-$L$ limits of $R_\xi$ and $U$ at the critical point, for
  cubic lattices with periodic boundary conditions). }
  \label{tablexy}
\end{table}

\begin{figure}[b]
  \includegraphics[width=0.95\columnwidth]{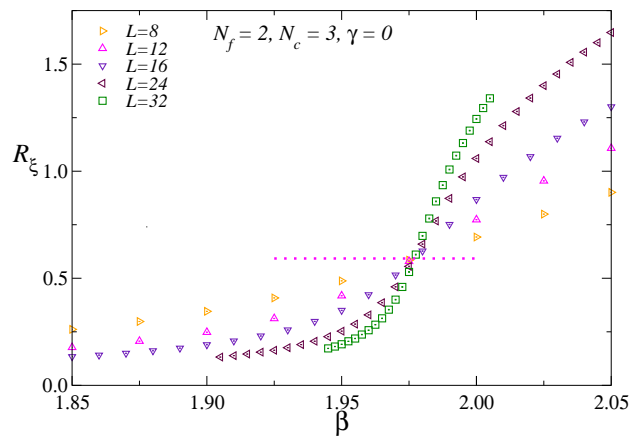}
  \caption{MC data of $R_\xi$ versus $\beta$ for the lattice SO(3)
    gauge model (\ref{hgauge}) with $N_f=2$ and $\gamma=0$.  The
    dotted line corresponds to $R_\xi = 0.5924$, which is the critical
    value for the $XY$ universality class, see Table~\ref{tablexy}.  }
\label{rxi-f2c3}
\end{figure}

\begin{figure}[t]
  \includegraphics[width=0.95\columnwidth]{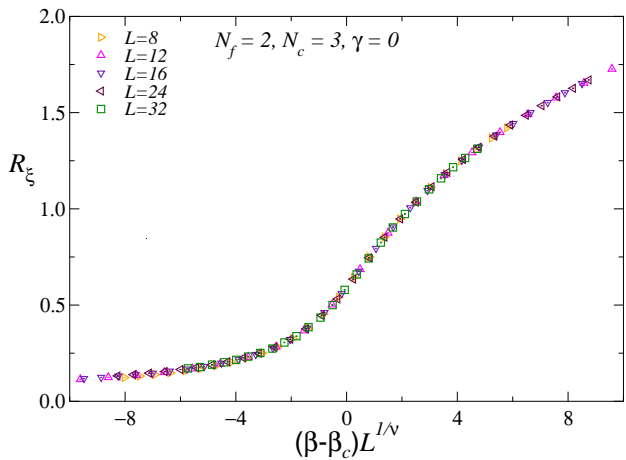}
  \caption{MC data of $R_\xi$ versus $(\beta-\beta_c)L^{1/\nu}$ for
    the lattice SO(3) gauge model (\ref{hgauge}) with $N_f=2$ and
    $\gamma=0$. We use $\nu = 0.6717$, the value for the $XY$
    universality class,  see Table~\ref{tablexy}. }
\label{rxi-f2c3-scaled}
\end{figure}

To begin with, we present results for $\gamma=0$. In
Fig.~\ref{rxi-f2c3} we show the estimates of $R_{\xi}$ for different
values of $L$ and $\beta$. The data sets for different values of $L$
clearly display a crossing point, which provides an estimate of the
critical point.  The data are consistent with the predicted $XY$
behavior.  Indeed, the data close to the crossing point nicely fit the
simple biased ansatz
\begin{equation}\label{unbfit}
R_\xi=R_\xi^* + a_1 X,\quad X=(\beta-\beta_c) L^{1/\nu},
\quad \nu=0.6717,
\end{equation}
where $\nu$ is the critical exponent of the $XY$ universality class,
see Table~\ref{tablexy}.  Using data within the self-consistent window
$R_\xi(\beta,L) \in [R_{\xi}^*(1 - \delta),R_{\xi}^*(1 + \delta)]$
with $\delta=0.2$, we obtain the estimate $\beta_c=
1.97690(7)$. The
collapse of the data in Fig.~\ref{rxi-f2c3-scaled}, where we report
the data of $R_\xi$ versus $X$, clearly shows the effectiveness of the
$XY$ biased fit.

\begin{figure}[!b]
  \includegraphics[width=0.95\columnwidth]{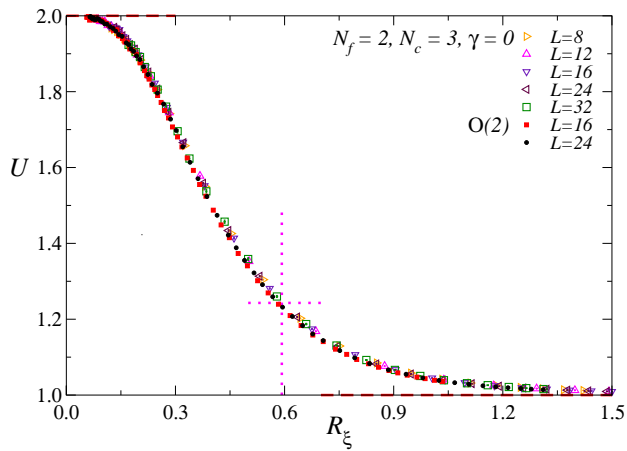}
  \caption{MC data of $U$ versus $R_\xi$ for the lattice SO(3) gauge
    model (\ref{hgauge}) with $N_f=2$ and $\gamma=0$ (data up to
    $L=32$) and for the $XY$ [O(2)] universality class (data up to
    $L=24$ for the standard nearest-neighbor $XY$ model).  The dotted
    lines correspond to the universal values $R^*_\xi$ and $U^*$ for
    the $XY$ universality class, see Table~\ref{tablexy}.  }
\label{nf2nc3}
\end{figure}

In Fig.~\ref{nf2nc3} we report results for the Binder parameter $U$
and the ratio $R_\xi=\xi/L$. The data of $U$ versus $R_\xi$ clearly
approach a single curve. which matches the corresponding curve
computed in the standard nearest-neighbor $XY$ model with action
(\ref{nvectorm}) (again we consider cubic lattices with periodic
boundary conditions).  This test, which does not require any tuning of
free parameters, provides the strongest evidence that the phase
transition in the gauge model belongs to the 3D $XY$ universality
class.  Note that scaling corrections are significantly smaller in the
gauge theory than in the standard discretizaton of the $XY$ model,
though they are hardly visible in Fig.~\ref{nf2nc3}.

\begin{figure}[t]
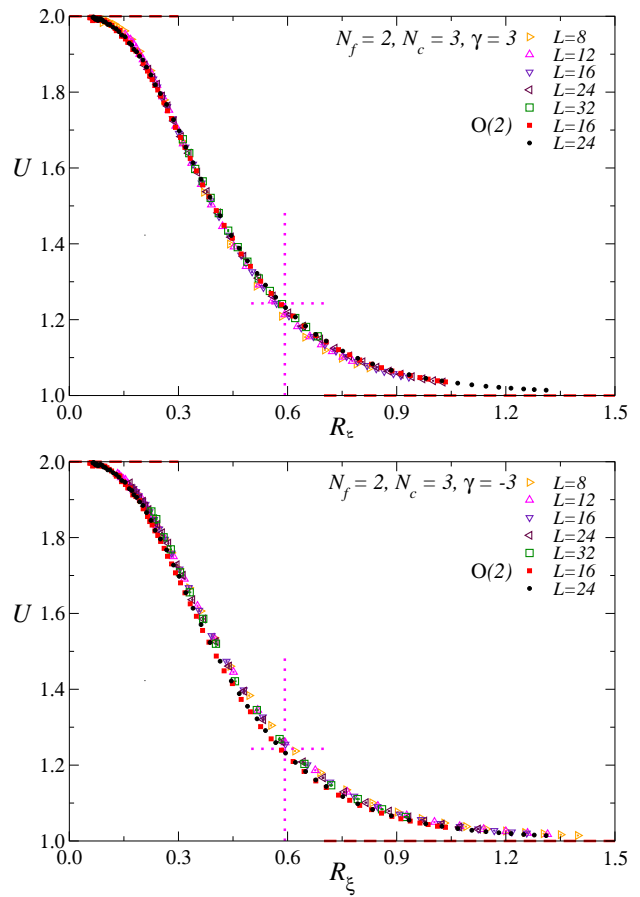

  \includegraphics[width=0.95\columnwidth]{urxi-f2c3-gamma3.eps}
  \includegraphics[width=0.95\columnwidth]{urxi-f2c3-gamma-3.eps}
  \caption{MC data of $U$ versus $R_\xi$ for the lattice SO(3) gauge
    model (\ref{hgauge}) with $N_f=2$ and $\gamma=\pm 3$ (up to
    $L=32$), and for the $XY$ [O(2)] universality class (data up to
    $L=24$ for the standard nearest-neighbor $XY$ model).  The dotted
    lines correspond to the universal values $R^*_\xi$ and $U^*$ for
    the $XY$ universality class, see Table~\ref{tablexy}.}
\label{nf2nc3-gamma}
\end{figure}

We have also checked that $XY$ behavior is also observed for other
values of $\gamma$; see Fig.~\ref{nf2nc3-gamma}, where we report
results for $\gamma=3$ and $-3$. This proves the irrelevance of
$\gamma$ along the transition line. Of course, a crossover is expected
in the limit $\gamma\to \infty$, Indeed, in this limit we should
observe an O($N$) critical behavior, with $N=N_f N_c = 6$ (results for
the O(6) universality class can be found in Ref.~\cite{AS-95}).

\subsection{The three-flavor lattice SO(3) gauge model}
\label{threefla}

For $N_f=3$ the LGW effective field theory predicts a first-order
phase transition for any number of colors. To verify the prediction,
we perform simulations for $\gamma=0$.

\begin{figure}[!tb]
  \includegraphics[width=0.95\columnwidth]{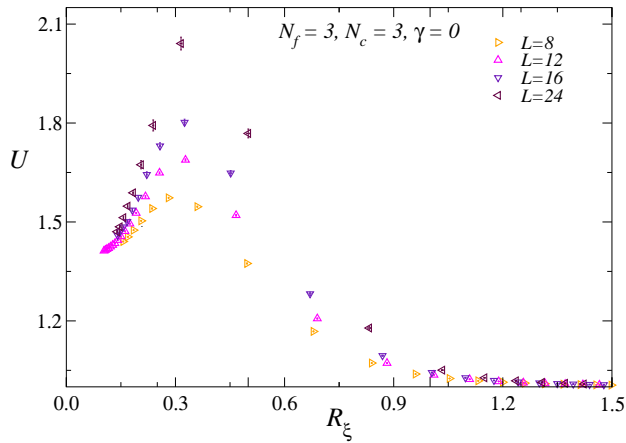}
  \caption{MC data of $U$ versus $R_\xi$ for the lattice SO(3) gauge
    model (\ref{hgauge}) with $N_f=3$ and $\gamma=0$. The absence of
    scaling indicates that the transition is not continuous, thus
     first order.}
\label{nf3nc3}
\end{figure}

\begin{figure}[!tb]
  \includegraphics[width=0.95\columnwidth]{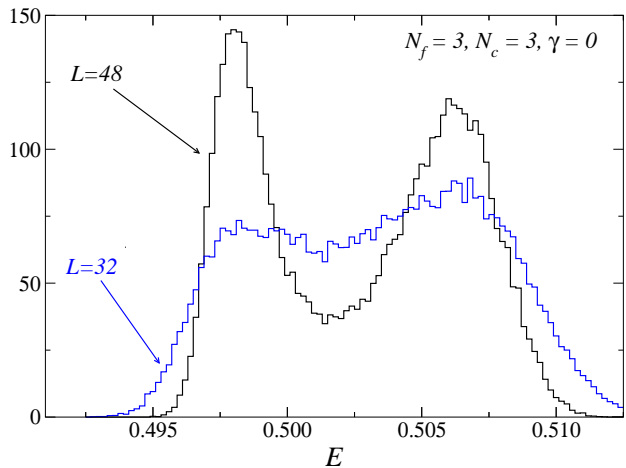}
  \caption{Energy distribution for the lattice SO(3) gauge model
    (\ref{hgauge}) with $N_f=3$ and $\gamma = 0$. 
    We report results for $L=32$ and $L = 48$ for $\beta$ values close 
    to the transition.
    The
    values of $\beta$ have been selected to obtain two maxima of
    approximately the same height.  }
\label{nf3nc3-double-peak}
\end{figure}

Some evidence in favor of a first-order transition is provided
by the analysis of the Binder parameter $U$. At a first-order
transition, the maximum $U_{\rm max}$ of $U$ behaves
as~\cite{CLB-86,VRSB-93} $U_{\rm max}\sim V$.  On the other hand, at a
continuous phase transition, $U$ is bounded as $L\to \infty$ and the
data of $U$ corresponding to different values of $R_{\xi}$ collapse
onto a common scaling curve as the volume is increased.  Therefore,
$U$ has a qualitatively different scaling behavior for first-order and
continuous transitions. In practice, a first-order transition can be
identified by verifying that $U_{\rm max}$ increases with $L$, without
the need of explicitly observing the linear behavior in the volume. A
second indication of a first-order transition is provided by the plot
of $U$ versus $R_{\xi}$. The absence of a data collapse is an early
indication of the first-order nature of the transition, as already
advocated in Ref.~\cite{PV-19}.  In Fig.~\ref{nf3nc3} we plot the
Binder parameter $U$ versus $R_{\xi}$.  The data, that are obtained at
values of $\beta$ close to the transition temperature $\beta_c \approx
1.7707$, do not show any scaling.  Moreover, $U$ displays a pronounced
peak, whose height increases with increasing volume. We take the
absence of scaling as an evidence that the transition is of first
order. 

The first-order transition is also clearly supported by the emergence
of a double peak structure in the distribution $P(E)$ of the energy
with increasing the lattice size around $\beta_c\approx 1.7707$.  This
is shown in Fig.~\ref{nf3nc3-double-peak} where the energy histograms
for $L=32$ and $L=48$ are compared.  Correspondingly the specific heat
$C_V$ defined in Eq.~\eqref{ecvdef} shows more and more pronounced
peaks with increasing $L$ (not shown).  However, the expected
asymptotic large-volume behaviors, such as $C_{V,{\rm max}}\sim V$ of
the maximum value $C_{V,{\rm max}}$ of $C_V$, are not clearly observed
yet, presumably requiring larger lattice sizes.

We have also considered the gauge-invariant two-point correlation
function of the local operator ${\rm det} \,\varphi_{\bm x}$ (note
that $\varphi^{af}_{\bm x}$ is the 3$\times$3 matrix), which may be
taken as an order parameter for the ${\mathbb Z}_2$ global symmetry
briefly discussed in Sec. \ref{sec:model}.  The correlation function
does not show any qualitative change across the transition. It is
always short-ranged, confirming that the ${\mathbb Z}_2$ global
symmetry is not broken and does not play any role at the transition.

In conclusion, the numerical results for $N_f=N_c=3$ provide a
convincing evidence that the transition is of first order for $\gamma
= 0$.  As it occurs for $N_f=2$, we conjecture that the nature of the
transition does not change in a large interval of values of $\gamma$
around $\gamma=0$.  In particular, we conjecture that the transition
is of first order for all positive finite values of $\gamma$. Note
that, for large $\gamma$, we expect significant crossover effects,
since the transition is continuous for $\gamma= \infty$ in the
universality class of the O(9) vector $\sigma$-model.

\section{Summary and conclusions}
\label{conclu}

In this paper we investigate the phase diagram of 3D multiflavor
lattice scalar theories in the presence of nonabelian SO($N_c$) gauge
interactions.  We consider the lattice scalar SO($N_c$) gauge theory
(\ref{hgauge}) with $N_f$ flavors, defined starting from a maximally
O($N$)-symmetric multicomponent scalar model ($N=N_f N_c$).  The
global O($N$) symmetry is partially gauged, obtaining a gauge model,
in which the fields belong to the coset $S^N$/SO($N_c$), where $S^N$
is the $N$-dimensional sphere.  Note that, for $N_c=2$, the action
(\ref{hgauge}) exactly maps onto that of the $N_f$-component lattice
Abelian-Higgs model characterized by a U(1) gauge symmetry, whose
phase diagram has been studied in Refs.~\cite{PV-19-2,PV-19}. We thus
focus on models with $N_c\ge 3$.

\begin{figure}[tbp]
\includegraphics*[scale=0.9]{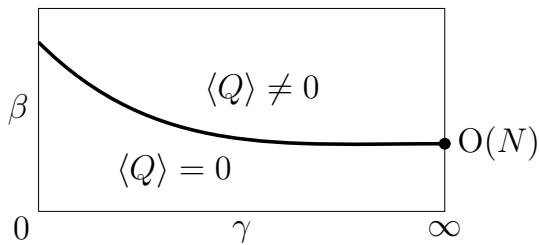}
\caption{Sketch of the phase diagram of the 3D lattice scalar
  SO($N_c$) gauge theory (\ref{hgauge}) with $N_f$ flavors and
  O($N_f$) global symmetry.  The transition line is continuous for
  $N_f=2$---it belongs to the $XY$ universality class for $N_c\ge 3$
  and to the O(3) universality class for $N_c=2$---and is of first
  order for $N_f\ge 3$.  We conjecture that its nature is the same for
  any finite $\gamma$.  The endpoint for $\gamma\to\infty$ is the
  O($N$) critical point ($N=N_cN_f$).  }
\label{phasediagram}
\end{figure}

For $N_f\ge 2$ the phase diagram is characterized by two phases: a
low-temperature phase in which the order parameter $Q_{\bm x}^{fg}$
defined in Eq.~(\ref{qdef}) condenses, and a high-temperature
disordered phase.  The two phases
are separated by a transition line, where the SO$(N_f)$ symmetry is
broken, as sketched in Fig.~\ref{phasediagram}.  The line ends at the
unstable O($N$) transition point with $N=N_cN_f$ for $\gamma \to
\infty$.  The gauge parameter $\gamma$, corresponding to the inverse
gauge coupling, does not play any particular role: the nature of the
transition is conjectured to be the same for any $\gamma$.  We have
numerically verified this conjecture for two values of $\gamma$.
Along the transition line only the correlations of the gauge-invariant
operator $Q_{\bm x}^{fg}$ are critical, while gauge modes are not
critical and only represent a background that gives rise to crossover
effects.

The nature of the finite-temperature transitions can be investigated
using different field-theoretical approaches. On one side, one can use
the effective LGW theory with Lagrangian~(\ref{SLGW}). In this
approach based on a gauge-invariant order parameter, only the global
symmetry group SO($N_f$) and the nature of the order parameter (a
rank-two symmetric real traceless tensor) play a role.  The gauge
degrees of freedom are absent in the effective model.  A second
approach is based on the continuum SO($N_c$) gauge theory, in which
the gauge fields are explicitly present.  
As it occurs for the lattice scalar chromodynamics characterized by an
SU($N_c$) gauge symmetry~\cite{BPV-19,BPV-20-3d}, the numerical
results agree with the LGW predictions.  The LGW framework provides
the correct description of the large-scale behavior of these systems,
predicting first-order transitions for $N_f=3$, and continuous
transitions for $N_f=2$, which belong to the $XY$ universality class
for any $N_c\ge 3$.  

The results for $N_f =2$ are in contradiction with the predictions of
the continuum gauge model (\ref{hgauge}): since no stable FP exists
for $N_f=2$, one would expect a first-order transition.  An analogous
contradiction was also observed in the case of scalar
chromodynamics~\cite{BPV-19}.  This apparent failure of the continuum
scalar gauge theory may suggest that it does not encode the relevant
modes at the transition.  Alternatively, the failure may be traced
back to the perturbative treatment around four dimensions, which does
not provide the correct description of the 3D behavior.  The 3D FP may
not be related to a four-dimensional FP, and therefore it escapes any
perturbative analysis in powers of $\varepsilon$. This has been also
observed in other physical systems; see, e.g.,
Refs.~\cite{MHS-02,CPPV-04}.  We finally recall that the two
field-theoretical approaches give different results also in the
large-$N_f$ limit. The LGW theory predicts a first-order transition
for any $N_f\ge 3$ due to the presence of the cubic term.  On the
other hand, continuous transitions are possible for large values of
$N_f$ according to the continuum scalar SO($N_c$) gauge theory, because of
the presence of a stable large-$N_f$ fixed
point~\cite{Hikami-80,PRV-01}.

\end{document}